\documentclass[12pt,a4paper]{article}
\usepackage[utf8]{inputenc}
\usepackage{amsmath}
\usepackage{amsfonts} 
\usepackage{mathtools} 
\usepackage{amssymb} 
\usepackage{subfigure} 
\usepackage{graphicx}
\usepackage{xcolor}
\usepackage{cite} 
\usepackage{mathrsfs}
\usepackage{hyperref}
\usepackage[normalem]{ulem}

\graphicspath{{Figs/}}

\newcommand{\abs}[1]{|#1|} 
\newcommand{\re}[1]{\text{Re}\left(#1\right)}
\newcommand{\im}[1]{\text{Im}\left(#1\right)}
\newcommand{\refEQ}[1]{eq.\,\eqref{#1}} 
\newcommand{\refEQS}[1]{eqs.\,\eqref{#1}} 

\newcommand{\Wq}[1]{{\rm W_{\rm #1}}}
\newcommand{\Wu}{\Wq{u}}
\newcommand{\Wd}{\Wq{d}}
\newcommand{\Wn}{\Wq{\nu}}
\newcommand{\Wl}{\Wq{\ell}}
\newcommand{\mq}[1]{m_{#1}}
\newcommand{\omq}[1]{\omega_{#1}}
\newcommand{\Xq}[1]{X_{#1}}
\newcommand{\Mq}[1]{\mathcal M_{#1}}
\newcommand{\mqd}{\mq{d}}\newcommand{\mqu}{\mq{u}}
\newcommand{\omqd}{\omq{d}}\newcommand{\omqu}{\omq{u}}
\newcommand{\Xqd}{\Xq{d}}\newcommand{\Xqu}{\Xq{u}}
\newcommand{\Mqd}{\Mq{d}}\newcommand{\Mqu}{\Mq{u}}

\newcommand{\CKM}{V}\newcommand{\CKMd}{\CKM^\dagger}
\newcommand{\V}[1]{{\CKM_{#1}^{\phantom{\ast}}}}
\newcommand{\Vc}[1]{{\CKM_{#1}^\ast}}
\newcommand{\PMNS}{U}\newcommand{\PMNSd}{U^\dagger}
\newcommand{\U}[1]{{\PMNS_{#1}^{\phantom{\ast}}}}

\newcommand{\wMU}{M_u^0}
\newcommand{\wMUd}{M_u^{0\dagger}}

\newcommand{\wMD}{M_d^0}
\newcommand{\wMDd}{M_d^{0\dagger}}

\newcounter{notas}
\graphicspath{{Figs/}}

\begin{document}

\hfill\begin{minipage}[r]{0.3\textwidth}\begin{flushright}  CFTP-17-010\\    IFIC-17-60 \end{flushright} \end{minipage}

\begin{center}

\vspace{0.50cm}

{\large \bf {Singlet Heavy Fermions as the Origin of B Anomalies in Flavour Changing Neutral Currents}}

\vspace{0.50cm}

Francisco J. Botella $^{a,}$\footnote{\texttt{Francisco.J.Botella@uv.es}},
Gustavo C. Branco $^{b,}$\footnote{\texttt{gbranco@tecnico.ulisboa.pt}},
Miguel Nebot $^{b,}$\footnote{\texttt{miguel.r.nebot.gomez@tecnico.ulisboa.pt}}
\end{center}

\vspace{0.50cm}
\begin{flushleft}
\emph{$^a$ Departament de F\`\i sica Te\`orica and Instituto de F\' \i sica Corpuscular (IFIC),\\
\quad Universitat de Val\`encia -- CSIC, E-46100 Valencia, Spain.}\\
\emph{$^b$ Departamento de F\'\i sica and Centro de F\' \i sica Te\' orica de Part\' \i culas (CFTP),\\
\quad Instituto Superior T\' ecnico (IST), U. de Lisboa (UL),\\ 
\quad Av. Rovisco Pais 1, P-1049-001 Lisboa, Portugal.} 
\end{flushleft}

\vspace{0.5cm}

\begin{abstract}
\noindent We show that a simple extension of the Standard Model involving the introduction of vector-like quarks and heavy neutrinos, provides an explanation of the so called B-anomalies in $b\to s\ell\bar\ell$ transitions. Vector-like quarks can explain, in the context of a discrete flavour symmetry, all the relevant characteristics of the Cabibbo-Kobayashi-Maskawa sector. It is in this framework that we study the requirements on the masses of the vector like quarks and the heavy neutrinos leading to viable models with sufficient deviations of lepton flavour universality and which simultaneously avoid too large Flavour Changing Neutral Current effects. Related predictions on $b\to d\ell\bar\ell$ and $s\to d\ell\bar\ell$ transitions are also analysed in detail.
\end{abstract}


\clearpage
\section{Introduction\label{SEC:intro}}
For the last few years there has been increasing evidence of what is generally known as ``B anomalies''. Two kinds of anomalies are emerging, related to the processes $b\to s\ell\bar\ell$ \cite{Aubert:2008bi,Lees:2012tva,Wei:2009zv,Wehle:2016yoi,Aaltonen:2011ja,Aaij:2013iag,Aaij:2013qta,Aaij:2015oid,Chatrchyan:2013cda,Khachatryan:2015isa,ATLAS:2017dlm,Sirunyan:2017dhj,Aaij:2014ora,Aaij:2017vbb}, and $b\to c\tau\nu$ \cite{Lees:2012xj,Lees:2013uzd,Bozek:2010xy,Huschle:2015rga,Aaij:2015yra,Aaij:2017uff}, \cite{Fajfer:2012vx,Becirevic:2012jf,Tanaka:2012nw,Freytsis:2015qca}. Here we are interested in the Flavour Changing Neutral Current (FCNC) anomaly related to the $b\to s\ell\bar\ell$ transition. The pioneer study \cite{Aaij:2013qta} of $B\to K^\ast\mu\bar\mu$ by the LHCb collaboration started \cite{Descotes-Genon:2013wba,Descotes-Genon:2013vna} with what now seems to indicate a systematic deficit, with respect to the Standard Model (SM) predictions, in several decay rates such as $B\to K^{(\ast)}\mu\bar\mu$ and $B_s\to\phi\mu\bar\mu$. LHCb has also published \cite{Aaij:2014ora,Aaij:2017vbb} results on
\begin{equation}\label{eq:RK:00}
R_{K^{(\ast)}}=\frac{\text{BR}(B\to K^{(\ast)}\mu\bar\mu)}{\text{BR}(B\to K^{(\ast)}e\bar e)}
\end{equation}
pointing to a deviation from the Lepton Flavour Universality (LFU) that holds in the SM. The Belle experiment has results supporting the LHCb findings \cite{Wehle:2016yoi}, the ATLAS and CMS collaborations also have results consistent with the aforementioned anomalies \cite{ATLAS:2017dlm,Sirunyan:2017dhj}.

Global fits to all these data have been done in the framework of an effective Hamiltonian approach where the New Physics (NP) contributions are included \cite{Bobeth:2012vn,Descotes-Genon:2013wba,Descotes-Genon:2013vna,Altmannshofer:2013foa,Horgan:2013pva,Hiller:2014yaa,Ghosh:2014awa,Altmannshofer:2014rta,Hurth:2016fbr,Capdevila:2016ivx,Serra:2016ivr,Altmannshofer:2017fio,Capdevila:2017bsm,Hiller:2017bzc,Ciuchini:2017mik,Hurth:2017sqw}, after excluding other possibilities and addressing in detail hadronic effects \cite{Khodjamirian:2010vf,Jager:2012uw,Descotes-Genon:2014uoa,Jager:2014rwa,Ciuchini:2015qxb,Straub:2015ica,Brass:2016efg,Capdevila:2017ert,Bobeth:2017vxj}, in the Wilson coefficients $C_i$ accompanying the operators
\begin{alignat}{3}\label{eq:Operators:00}
&\mathcal O_9^{[\mu]}=(\bar s\gamma^\nu\gamma_L b)(\bar \mu\gamma_\nu\mu),&&\quad \mathcal O_9^{[\mu]\prime}=(\bar s\gamma^\nu\gamma_R b)(\bar \mu\gamma_\nu\mu),\\
&\mathcal O_{10}^{[\mu]}=(\bar s\gamma^\nu\gamma_L b)(\bar \mu\gamma_\nu\gamma_5\mu),&&\quad \mathcal O_{10}^{[\mu]\prime}=(\bar s\gamma^\nu\gamma_R b)(\bar \mu\gamma_\nu\gamma_5\mu).
\end{alignat}
Among the viable NP scenarios, there are two broad categories depending on the amount and origin of LFU Violation (LFUV): (1) all the NP contribution is in the effective Hamiltonian for $\mathcal H_{\rm eff}(b\to s\mu\bar\mu)$, encoded in the coefficients $C_i^{[\mu]}$, and (2) the NP does also appear in the effective Hamiltonian $\mathcal H_{\rm eff}(b\to s e\bar e)$, in the coefficients $C_i^{[e]}$. NP analyses covering a wide range of scenarios, including Randall-Sundrum,  supersymmetric models, leptoquarks or $Z^\prime$ models, to name a few, have been put forward in recent times \cite{Altmannshofer:2013foa,Gauld:2013qba,Buras:2013qja,Datta:2013kja,Altmannshofer:2014rta,Hiller:2014yaa,Becirevic:2015asa,Celis:2015ara,Sahoo:2015qha,Cline:2015lqp,Boucenna:2016wpr,Crivellin:2016ejn,Hu:2016gpe,He:2017osj,Cline:2017ihf,Celis:2017doq,Bonilla:2017lsq,Chivukula:2017qsi,Megias:2017ove,Altmannshofer:2017yso,Megias:2017vdg}.

The model independent fits to $b\to s\ell\bar\ell$ and $b\to s\gamma$ data favour LFUV scenarios with NP entering mainly in $b\to s\mu\bar\mu$ transitions. We will concentrate on this favoured scenario, imposing that all NP contributions to $C_i^{[e]}$ are negligible.  When the fits are performed including NP contributions to the four operators in \refEQ{eq:Operators:00}, the most simple and favoured scenarios that arise can be classified according to where does the NP appear: (i) in $C_9^{[\mu]}$, (ii) in $C_9^{[\mu]}$ and $C_{10}^{[\mu]}$ with $C_9^{[\mu]}=-C_{10}^{[\mu]}$, (iii) in $C_9^{[\mu]}$ and $C_9^{[\mu]\prime}$ with $C_9^{[\mu]}=-C_9^{[\mu]\prime}$.
In the following we restrict ourselves to scenario (ii), where the origin of LFUV is NP in $b\to s\mu\bar\mu$ transitions with a $V-A$ structure. We thus omit in the rest of the paper the superscript $[\mu]$, with the understanding that $C_i$ means $C_i^{[\mu]}$. In order to specify the amount of NP that we are discussing, we recall that, at the scale $\mu_b=4.8$ GeV, the SM values of $C_{9,10}^{\rm SM}$ are $\sim 4.07$, $4.31$; the best fit in this scenario (ii) is obtained for $C_9\equiv C_{9}^{\rm SM}+\delta C_9$ with $\delta C_9\simeq -0.6$ .

\begin{figure}[h!tb]
\begin{center}
\subfigure[Box diagram.\label{fig:Box:00}]{\includegraphics[height=0.17\textwidth]{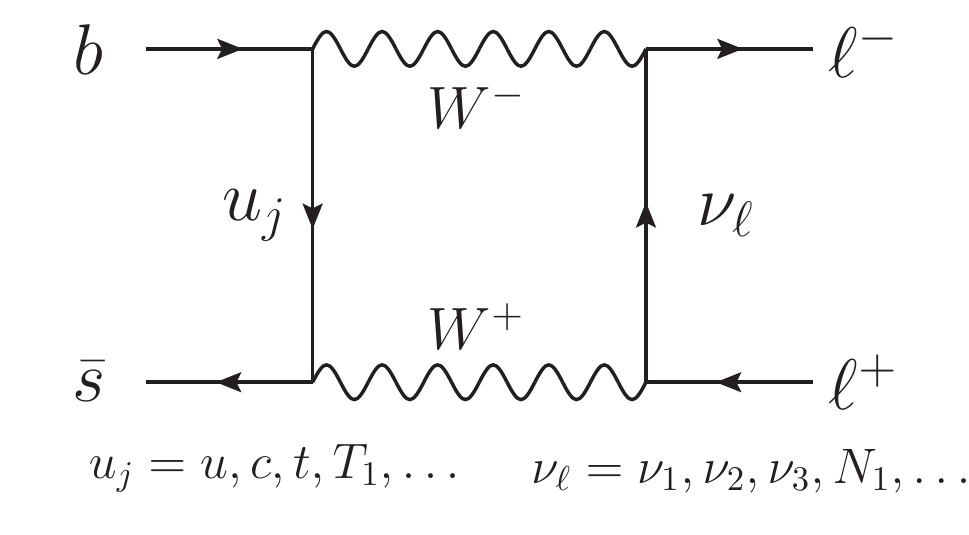}}\qquad
\subfigure[1PI Penguin diagrams.\label{fig:Penguin:00}]{\includegraphics[height=0.17\textwidth]{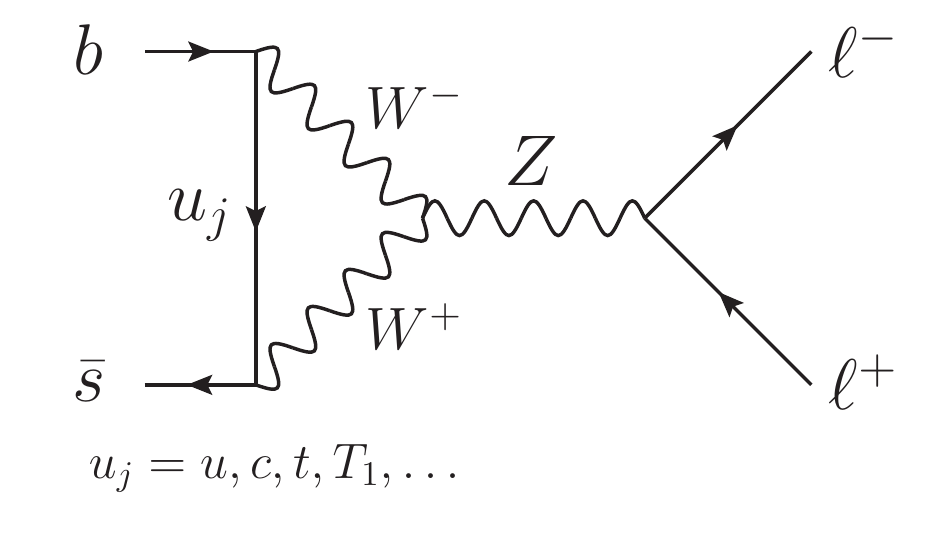}\quad\includegraphics[height=0.17\textwidth]{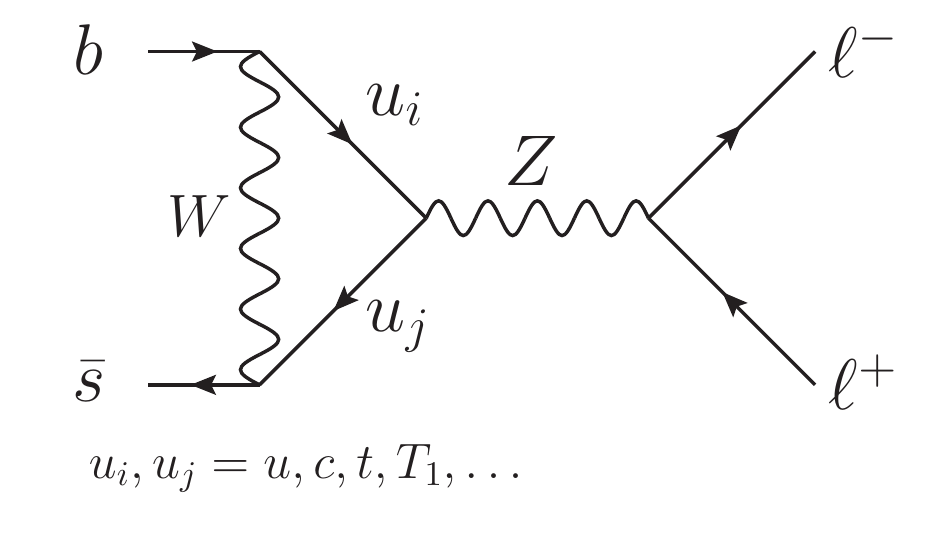}}
\caption{One loop contributions to $b\to s\ell\bar\ell$.\label{fig:Diagrams:bsll}}
\end{center}
\end{figure}

In this paper we study the possibility that the origin of these anomalies stems from the existence of new Heavy Fermions which mix with the Standard Model ones. In reference \cite{Botella:2016ibj} we have shown how vector-like quarks can explain, in the context of a discrete flavour symmetry, the origin of the light quark masses and of the small mixings of the Cabibbo-Kobayashi-Maskawa (CKM) matrix, providing a rationale for the almost decoupling of the third generation of quarks, with $\abs{\V{ub}}^2+\abs{\V{cb}}^2\sim 1.6\times 10^{-3}$. At the same time, one lesson in that same framework is that it is natural to have suppressed  (i.e. negligible) $Z$ and Higgs tree level mediated FCNCs among the SM quarks. Nevertheless, it was also pointed out in \cite{Botella:2016ibj} that one must be careful with the new one loop FCNC involving virtual heavy vector-like quarks. This general warning, taken into account in reference \cite{Botella:2016ibj}, can be translated into a relatively simple mechanism to generate New Physics contributions to $b\to s\ell\bar\ell$ transitions at the fifteen percent level. We are referring, for this process, to the contributions in the diagrams of figure \ref{fig:Diagrams:bsll}: the box \ref{fig:Box:00} and the $Z$ mediated penguin diagrams \ref{fig:Penguin:00}. With the SM contribution or even just introducing vector-like quarks, one cannot generate LFUV: one obtains the same corrections in $b\to s\mu\bar\mu$ and $b\to s e\bar e$. Concentrating on the box diagram, if, in addition, the $\nu_\ell$ has some small component of some heavy neutrino $N_i$, different for $\nu_\mu$ and $\nu_e$, then there is a chance to end up with corrections to $b\to s\ell\bar\ell$ with LFUV.

This is the avenue that we will explore (it has been partially considered in \cite{He:2017osj} including only the top quark contribution). 
The paper is organised as follows. In the next section, \ref{SEC:VLQ}, we specify the new fermions that are introduced. The corrections to the Wilson coefficients are analysed in section \ref{SEC:Analysis}. An example of this type of scenario which reproduces the $R_{K^{(\ast)}}$ anomalies is presented in section \ref{SEC:Example}. Correlated effects in other flavour processes are discussed in section \ref{SEC:Correlated}; we then conclude. In addition, appendix \ref{APP:EX} includes further details completing the example of section \ref{SEC:Example}, while in appendix \ref{APP:Univ} the relevant limits on deviations of $3\times 3$ unitarity of the leptonic mixing matrix are commented.

%
%

%

%

\section{Vector-like quarks\label{SEC:VLQ}}
We are interested in extensions of the SM where the Lagrangians for charged and neutral current interactions in the weak basis are given by
\begin{equation}\label{eq:LCC:00}
\mathscr L_W=-\frac{g}{\sqrt 2}\sum_{i=1}^3\left(\bar d^0_{Li}\gamma^\mu u^0_{Li}+\bar\ell_{Li}^0\gamma^\mu\nu^0_{Li}\right)W_\mu+\text{H.c.},
\end{equation}
and
\begin{multline}\label{eq:LNC:00}
\mathscr L_Z=-\frac{g}{\cos\theta_W}Z_\mu
\left[
\sum_{i=1}^3\left(\frac{1}{2}\bar u^0_{Li}\gamma^\mu u^0_{Li}-\frac{1}{2}\bar d^0_{Li}\gamma^\mu d^0_{Li}+\frac{1}{2}\bar \nu^0_{Li}\gamma^\mu \nu^0_{Li}-\frac{1}{2}\bar \ell^0_{Li}\gamma^\mu \ell^0_{Li}\right)\right.\\
\left.
-\sin^2\theta_W\left(\frac{2}{3}\sum_{i=1}^{3+n_u}\bar u^0_i\gamma^\mu u^0_i-\frac{1}{3}\sum_{i=1}^{3+n_d}\bar d^0_i\gamma^\mu d^0_i-\sum_{i=1}^{3}\bar\ell^0_i\gamma^\mu\ell^0_i\right)\right]\,,
\end{multline}
as in the SM, except for the electromagnetic current in the last line of \refEQ{eq:LNC:00}, since all the new fermions that we introduce are $SU(2)_L$ singlets. The new fields are $u^0_{Lk}$, $u^0_{Rk}$, $k=4,5,\ldots,3+n_u$ and similarly for $d^0$ and $\nu^0$ ($n_f$ is the number of new fermions of type $f$); to begin with, we do not need to introduce new charged leptons. In general, all the light fields mix with the heavy partners, which have the same colour and electric charges, but are singlets under $SU(2)_L$. The transformations of the fields into the mass eigenstate bases are given by
\begin{equation}
u^0_L=\begin{pmatrix}A_{uL}\\ B_{uL}\end{pmatrix}u_L,\quad d^0_L=\begin{pmatrix}A_{dL}\\ B_{dL}\end{pmatrix}d_L,\quad \nu^0_L=\begin{pmatrix}A_{\nu L}\\ B_{\nu L}\end{pmatrix}\nu_L,\quad \ell^0_L=\begin{pmatrix}A_{\ell L}\\ B_{\ell L}\end{pmatrix}\ell_L,
\end{equation}
where the matrices $\left(\begin{smallmatrix}A_{fL}\\ B_{fL}\end{smallmatrix}\right)$ are $(3+n_f)\times (3+n_f)$ unitary; $A_{fL}$ has $3$ rows and $3+n_f$ columns and, of course, $B_{fL}$ has $n_f$ rows and $3+n_f$ columns.
Note that a generalized GIM mechanism is encoded in the unitarity equations
\begin{align}
\begin{pmatrix}A_{uL}\\ B_{uL}\end{pmatrix}\begin{pmatrix}A_{uL}^\dagger & B_{uL}^\dagger\end{pmatrix}&=
\mathbf{1}_{(3+n_f)\times (3+n_f)}\Leftrightarrow 
\begin{pmatrix}A_{uL}A_{uL}^\dagger & A_{uL}A_{uL}^\dagger\\ B_{uL}A_{uL}^\dagger & B_{uL}B_{uL}^\dagger\end{pmatrix}=
\begin{pmatrix}\mathbf{1}_{3\times 3} & \mathbf{0}_{3\times n_f}\\ \mathbf{0}_{n_f\times 3} & \mathbf{1}_{n_f\times n_f}\end{pmatrix},\\
\begin{pmatrix}A_{uL}^\dagger & B_{uL}^\dagger\end{pmatrix}\begin{pmatrix}A_{uL}\\ B_{uL}\end{pmatrix}&=
\mathbf{1}_{(3+n_f)\times (3+n_f)}\Leftrightarrow 
A_{uL}^\dagger A_{uL}+ B_{uL}^\dagger B_{uL}=
\mathbf{1}_{(3+n_f)\times (3+n_f)}\,.
\end{align}
The first three components of $u_L$, $d_L$, $\nu_L$ and $\ell_L$ correspond to the known light matter degrees of freedom. In terms of physical fields, the Lagrangians of charged and neutral current interactions now read
\begin{equation}\label{eq:LCC:01}
\mathscr L_W=-\frac{g}{\sqrt 2}\left(\bar d_{L}\CKMd\gamma^\mu u_{L}+\bar\ell_{L}\PMNS\gamma^\mu\nu_{L}\right)W_\mu+\text{H.C.},
\end{equation}
and
\begin{multline}\label{eq:LNC:01}
\mathscr L_Z=-\frac{g}{\cos\theta_W}Z_\mu
\left[
\left(\frac{1}{2}\bar u_{L}\Wu\gamma^\mu u^0_{L}-\frac{1}{2}\bar d_{L}\Wd\gamma^\mu d_{L}+\frac{1}{2}\bar \nu_{L}\Wn\gamma^\mu \nu_{L}-\frac{1}{2}\bar \ell_{L}\Wl\gamma^\mu \ell_{L}\right)\right.\\
\left.
-\sin^2\theta_W\left(\frac{2}{3}\bar u\gamma^\mu u-\frac{1}{3}\bar d\gamma^\mu d-\bar\ell\gamma^\mu\ell\right)\right],
\end{multline}
with summations over all mass eigenstate indices understood. The generalized Cabibbo-Kobayashi-Maskawa matrix $\CKM$ is given by
\begin{equation}\label{eq:CKM:00}
\CKM\equiv A_{uL}^\dagger\, A_{dL}\,,
\end{equation}
and, obviously, it is not anymore a unitary matrix; it has in general $3+n_u$ rows and $3+n_d$ columns. The Pontecorvo-Maki-Nakagawa-Sakata matrix $\PMNS$ is
\begin{equation}\label{eq:PMNS:00}
\PMNS=A_{\ell L}^\dagger A_{\nu L}\,,
\end{equation}
which, again, is not unitary. Since we do not add heavy charged leptons, $\PMNS$ has $3$ rows and $3+n_\nu$ columns. The neutral current couplings in \refEQ{eq:LNC:01} are no longer diagonal, there are in general FCNC controlled by the matrices
\begin{equation}\label{eq:FCNC:W:00}
\Wu\equiv \CKM\CKMd,\quad \Wd=\CKMd\CKM,\quad \Wn=\PMNSd\PMNS,\quad \Wl=\mathbf{1}_{3\times 3}\,.
\end{equation}
The absence of FCNC in the charged lepton sector, $\Wl=\PMNSd\PMNS=\mathbf{1}_{3\times 3}$, is a direct consequence of the fact that we do not introduce heavy charged leptons. At this point, it is worthwhile mentioning that given the constraints on $|(\Wl)_{\mu\mu}-1|\leq 10^{-3}$ from $Z\to\mu^+\mu^-$ \cite{Fernandez-Martinez:2016lgt} and on $|(\Wd)_{bs}|\leq 10^{-5}$ from $B_s^0$--$\bar B_s^0$ mixing \cite{Barenboim:2001fd,Alok:2014yua}, Z mediated tree level FCNC are useless in order to explain the considered $b\to s$ anomalies.
It is also understood that besides the new $SU(2)_L$ left-handed singlets explicitely discussed, the corresponding right-handed singlets, as mentioned after \refEQ{eq:LNC:00}, are also introduced to fully define the theory.

\section{Anatomy of the Contributions to the Wilson Coefficients\label{SEC:Analysis}}
\subsection{The Standard Model\label{sSEC:SM:C910}}

The effective Lagrangian of interest for $b\to s\ell\bar\ell$ has the form
\begin{equation}\label{eq:Leff:bs:00}
\mathscr L_{\rm eff}=\frac{4G_F}{\sqrt 2}\V{tb}\Vc{ts}\frac{\alpha}{4\pi}\left(C_9\mathcal O_9+C_{10}\mathcal O_{10}\right)+\text{H.C.},
\end{equation}
with $\mathcal O_{9,10}$ in \refEQ{eq:Operators:00}; the coefficients $C_{9,10}$ arise from the diagrams in Figure \ref{fig:Diagrams:bsll}. In the SM and in the extensions with singlet vector-like quarks and also with heavy neutrinos,
\begin{equation}
C_9=-C_{10}\,.
\end{equation}
In the SM, $C_{9,10}=C_{9,10}^{\rm SM}$ comes from (a) the box diagram --Fig. \ref{fig:Box:00}-- with the three light quarks $u,c,t$, and the three massless neutrinos $\nu_{1,2,3}$ running in the loop, and (b) from the penguin diagrams --Fig. \ref{fig:Penguin:00}-- with the three light quarks. In an $R_\xi$ gauge, the box graph is gauge dependent, divergent in the unitary gauge, and the corresponding diagrams with would-be Goldstone bosons vanish with the light neutrino masses. Of course, gauge dependence cancels when the penguin diagrams are included. These diagrams were first calculated by Inami and Lim (IL) \cite{Inami:1980fz} separating the Landau gauge contribution and checking the cancellation of the gauge dependent parts among the box and the $Z$ penguins. They obtained, at the electroweak scale,
\begin{equation}\label{eq:C910SM:00}
C_{9}^{\rm SM}=-C_{10}^{\rm SM}=\frac{1}{\sin^2\theta_W}\sum_{\alpha=c,t}\frac{\Vc{\alpha s}\V{\alpha b}}{\Vc{ts}\V{tb}}Y(x_\alpha)\sim \frac{Y(x_t)}{\sin^2\theta_W},
\end{equation}
with $x_t=m_t^2/M_W^2$ and the IL function
\begin{equation}\label{eq:IL:Y}
Y(x)=\frac{x}{8}\left(\frac{x-4}{x-1}+\frac{3x\ln x}{(x-1)^2}\right)\,.
\end{equation}

\subsection{Singlet Vector-like Quarks\label{sSEC:VL:C910}}
As explained before, we consider the scenario in \cite{Botella:2016ibj} as our singlet vector-like quark extension. In particular, we introduce $n$ up and $n$ down singlets, labelled $T_{1L},T_{1R}$, $\ldots$, $T_{nL},T_{nR}$ and $D_{1L},D_{1R}$, $\ldots$, $D_{nL},D_{nR}$ respectively (in this obvious notation, $T_{j}=u_{3+j}$ and $D_{j}=d_{3+j}$ in \refEQ{eq:LCC:01}--\eqref{eq:LNC:01}). For the purpose of the present work, it is sufficient to introduce one vectorial quark in order to have, inside the loops in Figure \ref{fig:Diagrams:bsll}, a quark heavier than the top. However, in order to explain light quark masses and small mixings in a symmetry controlled context, we use the complete scenario introduced in \cite{Botella:2016ibj}. Without considering the neutrino extension yet, let us consider again the diagrams in Figure \ref{fig:Diagrams:bsll}. Taking into account that the IL result is gauge invariant for any number of generations, and that the box diagram has the same form as in the SM enlarged with $n$ additional generations, the best way to present the calculation in this extension is to split the Wilson coefficient $C_9^{\rm VLQ}$ in a SM-like contribution for $3+n$ generations $C_9^{\rm SM}(3+n)$ and the remaining piece $\Delta C_9^{\rm HQ}$ (``HQ'' for Heavy Quark):
\begin{equation}
C_9^{\rm VLQ}=C_9^{\rm SM}(3+n)+\Delta C_9^{\rm HQ}\,.
\end{equation}
The SM-like contribution with $3+n$ generations is
\begin{equation}\label{eq:C9SMn3:00}
C_9^{\rm SM}(3+n)=\frac{1}{\sin^2\theta_W}\sum_{\alpha=c,t,T_1,\ldots,T_n}\frac{\Vc{\alpha s}\V{\alpha b}}{\Vc{ts}\V{tb}}Y(x_\alpha)\,.
\end{equation}
The question now is to find out the remaining piece $\Delta C_9^{\rm HQ}$. As already mentioned, $C_9^{\rm SM}(3+n)$ includes all the contributions with SM couplings of the $Z$ to fermions. Neglecting the contributions with tree level flavour changing couplings of the $Z$ to down quarks, much constrained by $B_s$ meson mixing, it is clear that the only contributions not properly included in $C_9^{\rm SM}(3+n)$ appear in the right diagram of Fig. \ref{fig:Penguin:00}: the $\Delta C_9^{\rm HQ}$ piece comes exactly from that penguin diagram with the substitution $\Wu\mapsto \Wu-\mathbf{1}$. It can be shown that this contribution is gauge invariant on its own, and previous calculations can be found in \cite{Nardi:1995fq,Vysotsky:2006fx,Kopnin:2008ca,Picek:2008dd}. Explicitly,
\begin{equation}
\Delta C_9^{\rm HQ}=\frac{1}{\sin^2\theta_W}\sum_{\alpha=c,t,T_1,\ldots,T_n}\frac{\Vc{\alpha s}(\Wu-\mathbf{1})_{\alpha\beta}\V{\beta b}}{\Vc{ts}\V{tb}}N(x_\alpha,x_\beta),\qquad x_\alpha=\frac{m_\alpha^2}{M_W^2},
\end{equation}
where
\begin{equation}\label{eq:loopN:00}
N(x_\alpha,x_\beta)=\frac{xy}{8(x-y)}(\ln x-\ln y)\,.
\end{equation}
Vector-like quarks do not violate decoupling requirements \cite{Appelquist:1974tg,Picek:2008dd}, but the IL function $Y(x)$ grows with $x$: it can be checked that the addition of $\Delta C_9^{\rm HQ}$ to $C_9^{\rm SM}(3+n)$ enforces decoupling: noticing that in the decoupling limit \cite{Botella:2016ibj}
\begin{equation}\label{eq:decoupling:00}
\Wu-\mathbf{1}\to\begin{pmatrix} \mathbf{0}_{3\times 3} & \mathbf{0}_{3\times n}\\ \mathbf{0}_{n\times 3} & -\mathbf{1}_{n\times n}\end{pmatrix}
\end{equation}
and
\begin{equation}\label{eq:loopN:01}
N(x,x)=\frac{x}{8}\,,
\end{equation}
the contribution of the heavy quark $T_i$ to $C_9^{\rm SM}(3+n)+\Delta C_9^{\rm HQ}$ is of the form
\begin{equation}\label{eq:decoupling:01}
Y(x_{T_i})-\frac{x_{T_i}}{8}=\frac{x_{T_i}}{8}\left(\frac{-3}{x_{T_i}-1}+\frac{3x_{T_i}\ln x_{T_i}}{(x_{T_i}-1)^2}\right)\xrightarrow{x_{T_i}\to\infty}-\frac{3}{8}+\frac{3}{8}\ln x_{T_i}
\end{equation}
which does not grow as $x_{T_i}$ anymore enforcing decoupling. It is to be stressed, however, that the Wilson coefficient $C_9^{\rm VLQ}$ is still Lepton Flavour Universal. It is clear that to get a natural LFUV contribution we do not want this VLQ contribution to be important since it is LFU. This implies that our vectorial quark should be heavy enough to reduce this contribution. Our next step is to introduce heavy neutrinos to induce LFUV.

\subsection{Singlet Vector-like Quarks and Heavy Neutrinos\label{sSEC:NVL:C910}}
In order to achieve LFUV, we will keep the neutrino sector as general as possible. That is, we will assume that running in the box diagram of Fig. \ref{fig:Box:00} there are three essentially massless neutrinos $\nu_1,\nu_2,\nu_3$ and in addition heavy neutrinos $N_j$. The contribution to $b\to s\ell\bar\ell$ is proportional to $\abs{\U{\ell \nu_i}}^2$ for each light neutrino, and to $\abs{\U{\ell N_j}}^2$ for each heavy one, with
\begin{equation}\label{eq:PMNS:Unit:00}
\sum_{i=1}^3\abs{\U{\ell \nu_i}}^2=1-\sum_{j=4}^{n_N}\abs{\U{\ell N_j}}^2\,.
\end{equation}
The contribution to $C_9^{\rm SM}$ and to $C_9^{\rm VLQ}$ mediated by light neutrinos is proportional to $\sum_{\nu_i=1}^3\abs{\U{\ell \nu_i}}^2$, while the one mediated by the heavy neutrinos $N_j$ in the box (Fig. \ref{fig:Box:00}) will be proportional to $\sum_{j=4}^{n_N}\abs{\U{\ell N_j}}^2 q^2/(q^2-m_{N_j}^2)$, giving
\begin{multline}\label{eq:HeavyNeutrinoBox:00}
\sum_{\nu_i=1}^3\abs{\U{\ell \nu_i}}^2+\sum_{j=4}^{n_N}\abs{\U{\ell N_j}}^2 \frac{q^2}{q^2-m_{N_j}^2}=1+\sum_{j=4}^{n_N}\abs{\U{\ell N_j}}^2 \left(\frac{q^2}{q^2-m_{N_j}^2}-1\right)=\\ 
1+\sum_{j=4}^{n_N}\abs{\U{\ell N_j}}^2\frac{m_{N_j}^2}{q^2-m_{N_j}^2}
\end{multline}
from the neutrino propagators (the first term, ``1'', is just the standard contribution). As we know, this piece is gauge dependent, but for the heavy neutrinos the box diagrams with would-be Goldstone bosons do not vanish anymore, reestablishing gauge invariance. The complete result for the model including vector-like quarks and enlarged with heavy neutrinos (labelled ``VLN'' in the following) is
\begin{equation}\label{eq:FullC9:00}
C_9^{\rm VLN}=C_9^{\rm SM}(3+n)+\Delta C_9^{\rm HQ}+\Delta C_9^{\rm HN}\,,
\end{equation}
where the last term $\Delta C_9^{\rm HN}$ is the heavy neutrino contribution which can violate Lepton Flavour Universality:
\begin{equation}\label{eq:C9HN:00}
\Delta C_9^{\rm HN}=\frac{1}{4\sin^2\theta_W}\sum_{\alpha=c,t,T_1,\ldots,T_n}\frac{\Vc{\alpha s}\V{\alpha b}}{\Vc{ts}\V{tb}}\sum_{j=4}^{n_N}\abs{\U{\ell N_j}}^2 G(x_{\alpha j},x_{Wj}),
\end{equation}
with $x_{\alpha j}=(m_\alpha/m_{N_j})^2$, $x_{Wj}=(M_W/m_{N_j})^2$ and 
\begin{equation}\label{eq:Gloop:00}
G(x,y)=\frac{3x(1-x-y+xy)}{4(x-y)(1-x)(1-y)^2}+\frac{3x(2x-y-xy)\ln y}{4(x-y)^2(1-y)^2}+\frac{(x^2+4y^2-8xy)x\ln x}{4y(x-y)^2(1-x)}\,.
\end{equation}
$G(x,y)$ is in agreement with \cite{He:2017osj}, where 
\begin{equation}\label{eq:GEloop:00}
E((m_\alpha/M_W)^2,(m_{N_j}/M_W)^2)=-G((m_\alpha/m_{N_j})^2,(M_W/m_{N_j})^2)
\end{equation} is used instead.

%

\section{Heavy Neutrinos and no Vector-like Quarks\label{SEC:4}}
\begin{figure}[h!tb]
\begin{center}
\subfigure[$-G$ vs. $m_N$ for fixed $m_T$ values.\label{fig:BoxHN:1}]{\includegraphics[width=0.4\textwidth]{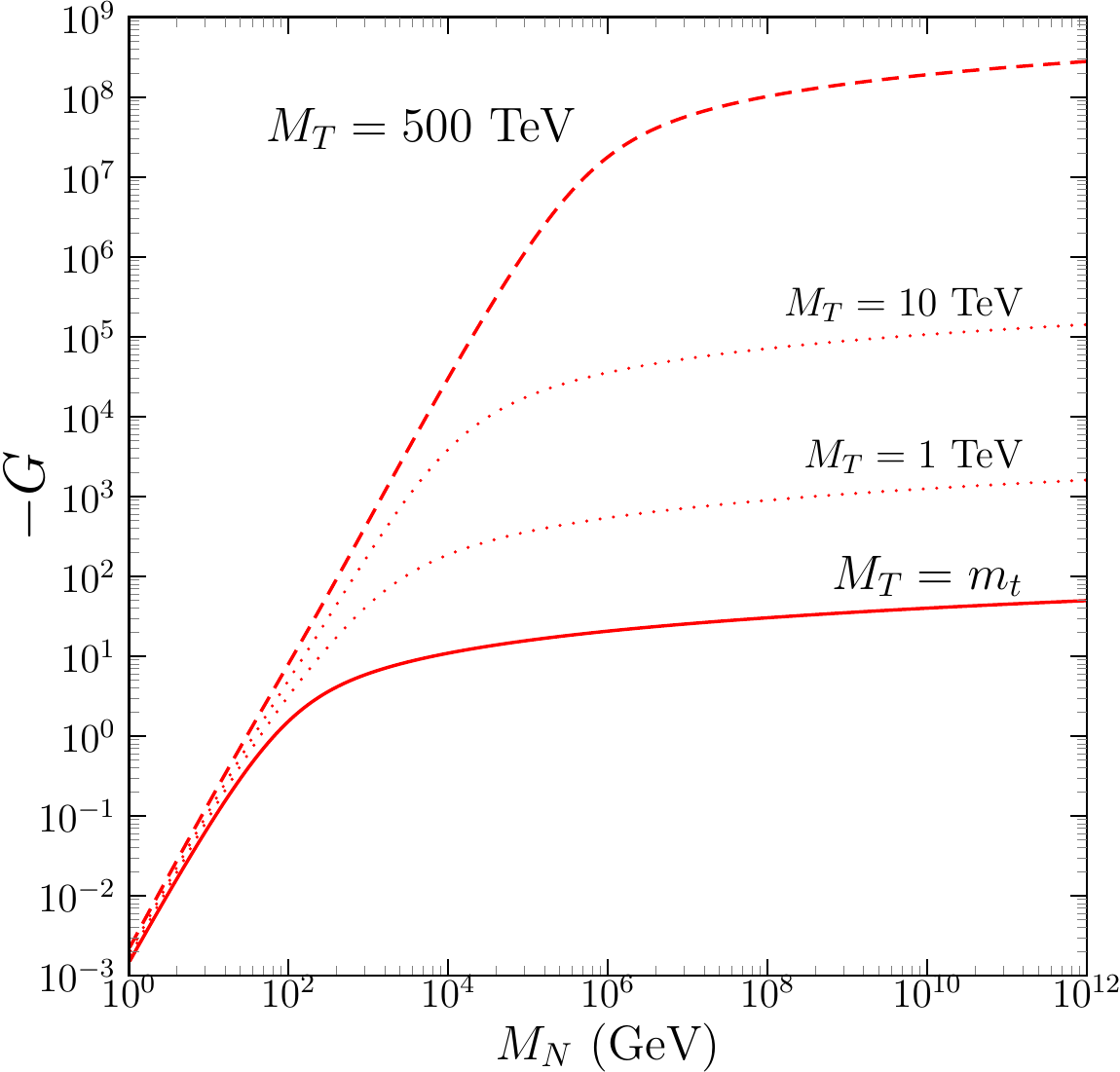}}\qquad
\subfigure[$-G$ contours in $(m_T,m_N)$.\label{fig:BoxHN:2}]{\includegraphics[width=0.39\textwidth]{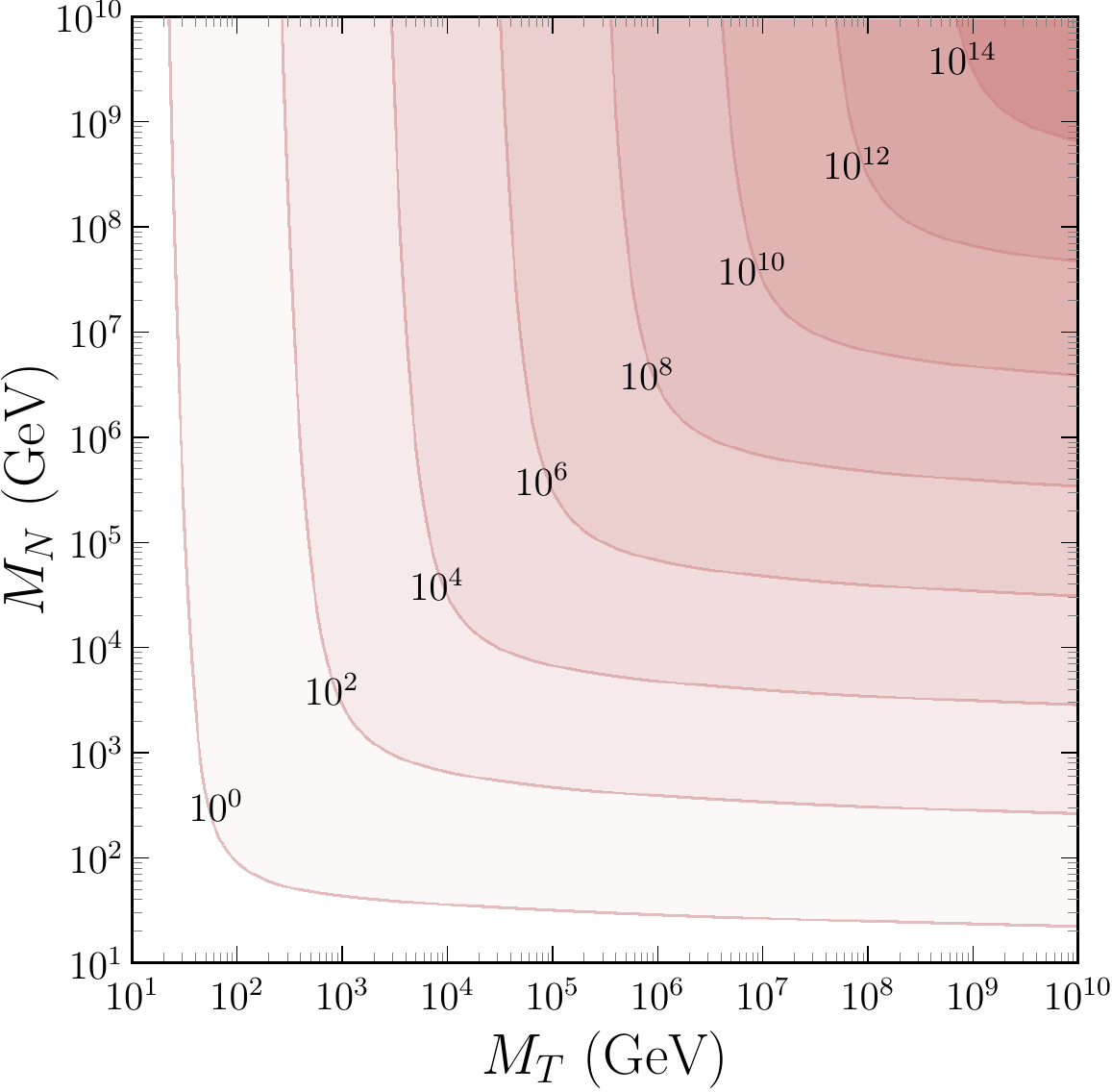}}
\caption{Box function $G$ with heavy neutrinos in \refEQ{eq:Gloop:00}.\label{fig:BoxHN}}
\end{center}
\end{figure}

With the results from the previous section, we can now address the anomalies quantitatively. Since LFUV are related to the presence of heavy neutrinos, before addressing the anomalies in the complete scenario with both vector-like quarks and heavy neutrinos, we first analyse briefly what is the situation when heavy neutrinos are added, but no vector-like quarks.
In that case, one has
\begin{equation}\label{eq:C9noVLq:00}
C_9^{\rm VLN}\to C_9^{\rm N}=C_9^{\rm SM}(3+0)+\Delta C_9^{\rm HQ}+\Delta C_9^{\rm SM-HN}=C_9^{\rm SM}+\Delta C_9^{\rm SM-HN}
\end{equation}
where $\Delta C_9^{\rm HN}\to\Delta C_9^{\rm SM-HN}$ when the quark sector is simply the SM one, that is
\begin{equation}\label{eq_C9noVLq:01}
\Delta C_9^{\rm SM-HN}=\frac{1}{4\sin^2\theta_W}\sum_{j=4}^{n_N}\abs{\U{\ell N_j}}^2 G\left(\frac{m_t^2}{m_{N_j}^2},\frac{M_W^2}{m_{N_j}^2}\right)\,,
\end{equation}
omitting negligible charm contributions. It has to be stressed that the function $G$ is negative in all the parameter space, and thus in this scenario one obtains naturally a negative contribution to $C_9$ from heavy neutrinos which mix with the light ones. We plot $-G$ in Figure \ref{fig:BoxHN} as a function of a heavy neutrino mass $M_N$ and an up type quark mass $M_T$.
We can read in Fig. \ref{fig:BoxHN:1} that for the top quark mediated contribution, $-G(x,y)< 10^2$ even if $M_N$ arrives to values as large as $10^{12}$ GeV. As discussed later, following universality constraints, $\abs{\U{\ell N_j}}^2\leq 10^{-3}$ and thus it is very difficult to reach values $\Delta C_9^{\rm SM-HN}\sim -0.1$ only with the top quark contribution. Unless they are extremely heavy, it turns out that just the presence of heavy neutrinos cannot be the origin of the B anomalies, including the LFUV $R_{K^{(\ast)}}$ anomalies.\\
From figure \ref{fig:BoxHN:1}, one can see that for a large $m_T$ there is region where $-G$ grows linearly with $m_N$ until $m_N\sim m_T$ (see for example the dashed line for $M_T=500$ TeV). With $m_N$ towards the end of this rapidly growing region, we can optimise the New Physics effect.


\section{Reproducing the Anomalies with Heavy Neutrinos and Vector-like Quarks\label{SEC:Example}}
It is now clear that the presence of heavy neutrinos or of heavy vector-like quarks, but not both simultaneously, is not sufficient to reproduce the B anomalies. In this section we address the complete scenario with both heavy neutrinos and heavy vector-like quarks, and present an example successful in that scope. 
\subsection{Quark sector\label{sSEC:QuarkSector}}
As discussed in section \ref{SEC:VLQ}, the quark sector of the model contains three up vector-like singlets $T_{jL}^0$, $T_{jR}^0$, $j=1,2,3$, and three down vector-like singlets $D_{jL}^0$, $D_{jR}^0$, $j=1,2,3$. In principle, in order to address the B anomalies, we do not need to add that many vectorial quarks. However, since it is interesting to have a realistic model with motivations going beyond accounting for the anomalies, we consider the complete scenario, which explains light quark masses and mixings in a symmetry based approach \cite{Botella:2016ibj}. In this context, the relevant part of the Lagrangian is the mass matrix after spontaneous electroweak symmetry breaking:
\begin{equation}\label{eq:QLMass:00}
\mathscr L_{\rm M}=
-\begin{pmatrix}\bar d_L^0& \bar D_L^0\end{pmatrix}\,\wMD\,\begin{pmatrix}d_R^0\\ D_R^0\end{pmatrix}
-\begin{pmatrix}\bar u_L^0& \bar T_L^0\end{pmatrix}\,\wMU\,\begin{pmatrix}u_R^0\\ T_R^0\end{pmatrix},
\end{equation}
with the $6\times 6$ mass matrices $\wMD$ and $\wMU$ further divided in $3\times 3$ submatrices
\begin{equation}\label{eq:MassBlocks:00}
\wMD\equiv
\begin{pmatrix}\mqd & \omqd \\ \Xqd & \Mqd\end{pmatrix}.\qquad
\wMU\equiv
\begin{pmatrix}\mqu & \omqu \\ \Xqu & \Mqu\end{pmatrix}\,.
\end{equation}
$\mq{q}$ and $\omq{q}$ are $\Delta I=1/2$ mass terms (of the electroweak scale order) while $\Xq{q}$ and $\Mq{q}$ are $\Delta I=0$ bare mass terms of the scale of the heavy vectorial quark masses. The explicit mass matrices are shown in appendix \ref{APP:EX}. In any case, in order to check consistency of all the relevant flavour data and in order to analyse all possible correlated effects (if any), we need a very detailed model. The masses of the quarks, in GeV, are:
\begin{equation}\label{eq:QMasses:00}
\begin{pmatrix}
m_d \\ m_s \\ m_b \\ m_{D_1} \\ m_{D_2} \\ m_{D_3}
\end{pmatrix}=
\begin{pmatrix}
0.0029 \\ 0.055 \\ 2.9 \\ 10012.5 \\ 11011.5 \\ 12010.4
\end{pmatrix}\,,\qquad
\begin{pmatrix}
m_u \\ m_c \\ m_t \\ m_{T_1} \\ m_{T_2} \\ m_{T_3}
\end{pmatrix}=
\begin{pmatrix}
0.0013 \\ 0.68 \\ 171.5 \\ 502494 \\ 503154 \\ 503891
\end{pmatrix}\,.
\end{equation}
The new down quarks have masses in the 10 TeV range while the up quarks are in the 500 TeV range. From the diagonalisation of the mass matrices in \refEQ{eq:QLMass:00}, (\refEQS{eq:MassM:m:00}--\eqref{eq:MassM:X:00} in appendix \ref{APP:EX}), we obtain the non-unitary $6\times 6$ CKM matrix $\CKM$. Its moduli are given by
\begin{equation}\label{eq:CKM:Example:00}
{\small 
\abs{\CKM}=
\begin{pmatrix}
0.97427 		&0.22534		&0.00350		&3.7\cdot 10^{-5}	&6.4\cdot 10^{-6}	&7.2\cdot 10^{-7}\\
0.22521			&0.97348		&0.04099		&5.8\cdot 10^{-4}	&2.0\cdot 10^{-4}	&2.3\cdot 10^{-5}\\
0.00834			&0.04028		&0.99914		&1.7\cdot 10^{-5}	&5.1\cdot 10^{-3}	&1.4\cdot 10^{-4}\\
1.3\cdot 10^{-4}	&6.6\cdot 10^{-4}	&6.2\cdot 10^{-4}	&3.8\cdot 10^{-7}	&3.2\cdot 10^{-6}	&1.7\cdot 10^{-7}\\
2.0\cdot 10^{-6}	&1.2\cdot 10^{-5}	&2.3\cdot 10^{-4}	&1.0\cdot 10^{-8}	&1.2\cdot 10^{-6}	&3.2\cdot 10^{-8}\\
1.6\cdot 10^{-7}	&8.0\cdot 10^{-7}	&4.0\cdot 10^{-6}	&4.0\cdot 10^{-10}	&2.1\cdot 10^{-8}	&6.0\cdot 10^{-10}
\end{pmatrix},
}
\end{equation}
while the arguments of its elements (in a convenient phase convention \cite{Branco:1999fs}) are 
\begin{equation}
{\small 
\arg(\CKM)=
\begin{pmatrix}
0	&5.885\cdot 10^{-4}	&-1.119		&0.361		&-1.039		&1.401\\
\pi	&0			&0		&0.963		&0.331		&1.129\\	
-0.377	&\pi+0.018		&0		&-0.862		&0.113		&-2.013\\	
-1.896	&1.219			&-1.923		&2.210		&-1.818		&2.348\\	
-0.825	&2.308			&-1.976		&-2.444		&-1.862		&2.305\\
1.174	&-1.921			&1.219		&-0.794		&1.331		&-0.794
\end{pmatrix}.
}
\end{equation}
Finally, for the intensity of the FCNC, we give the moduli of the matrices $\Wu=\CKM\CKMd$ and $\Wd=\CKMd\CKM$,
\begin{equation}\label{eq:Wu:Example:00}
{\small 
\abs{\Wu}=
\begin{pmatrix}
1 			&1.3\cdot 10^{-8}	&1.3\cdot 10^{-8}	&2.1\cdot 10^{-5}	&2.1\cdot 10^{-9}	&2.1\cdot 10^{-8}\\
1.3\cdot 10^{-8}	&1			&4.2\cdot 10^{-7}	&6.5\cdot 10^{-4}	&1.2\cdot 10^{-5}	&6.5\cdot 10^{-7}\\
1.3\cdot 10^{-8}	&4.2\cdot 10^{-7}	&1			&6.5\cdot 10^{-4}	&2.3\cdot 10^{-4}	&4.1\cdot 10^{-6}\\
2.1\cdot 10^{-5}	&6.5\cdot 10^{-4}	&6.5\cdot 10^{-4}	&8.4\cdot 10^{-7}	&1.4\cdot 10^{-7}	&3.1\cdot 10^{-9}\\
2.1\cdot 10^{-9}	&1.2\cdot 10^{-5}	&2.3\cdot 10^{-4}	&1.4\cdot 10^{-7}	&5.1\cdot 10^{-8}	&9.1\cdot 10^{-10}\\
2.1\cdot 10^{-8}	&6.5\cdot 10^{-7}	&4.1\cdot 10^{-6}	&3.1\cdot 10^{-9}	&9.1\cdot 10^{-10}	&1.7\cdot 10^{-11}
\end{pmatrix},
}
\end{equation}
\begin{equation}\label{eq:Wd:Example:00}
{\small 
\abs{\Wd}=
\begin{pmatrix}
1 			&5.9\cdot 10^{-8}	&3.4\cdot 10^{-9}	&1.0\cdot 10^{-4}	&7.4\cdot 10^{-7}	&6.0\cdot 10^{-6}\\
5.9\cdot 10^{-8}	&1			&2.5\cdot 10^{-7}	&5.7\cdot 10^{-4}	&4.8\cdot 10^{-5}	&2.8\cdot 10^{-5}\\
3.4\cdot 10^{-9}	&2.5\cdot 10^{-7}	&0.999973 		&2.6\cdot 10^{-5}	&1.7\cdot 10^{-16}	&2.2\cdot 10^{-18}\\
1.0\cdot 10^{-4}	&5.7\cdot 10^{-4}	&2.6\cdot 10^{-5}	&3.4\cdot 10^{-7}	&1.4\cdot 10^{-7}	&1.4\cdot 10^{-8}\\
7.4\cdot 10^{-7}	&4.8\cdot 10^{-5}	&1.7\cdot 10^{-16}	&1.4\cdot 10^{-7}	&2.6\cdot 10^{-5}	&7.3\cdot 10^{-7}\\
6.0\cdot 10^{-6}	&2.8\cdot 10^{-5}	&2.2\cdot 10^{-18}	&1.4\cdot 10^{-8}	&7.3\cdot 10^{-7}	&2.1\cdot 10^{-8}
\end{pmatrix}.
}
\end{equation}
In the light sector, it is well known that the potentially most dangerous FCNC are those mediated by $Z$ exchange (compared to Higgs exchange), and therefore controlled by the off-diagonal elements of the upper left $3\times 3$ submatrices in $\Wd$ and $\Wu$. Their contribution to meson mixing is very small, being proportional to $([\Wd]_{ij})^2$, which are of order $10^{-14}$ or smaller. For a $d_i\to d_j$ transition, the contributions to $C_{9,10}$ from tree level $Z$ exchange are \cite{Barenboim:2001fd,Alok:2014yua}
\begin{equation}\label{eq:treeZFCNC:00}
\delta C_{9}^{\text{tree}}=\frac{\pi[\Wd]_{d_jd_i}}{\alpha \Vc{td_j}\V{td_i}}(4\sin^2\theta_W-1),\qquad
\delta C_{10}^{\text{tree}}=\frac{\pi[\Wd]_{d_jd_i}}{\alpha \Vc{td_j}\V{td_i}}\,.
\end{equation}
For the $b\to s$ transition in particular, with \refEQ{eq:Wd:Example:00},
\begin{equation}
Z\text{ tree level FCNC:  }\abs{\delta C_{9}^{\text{tree}}}\simeq 2.6\times 10^{-3}(1-4\sin^2\theta_W),\quad\abs{\delta C_{10}^{\text{tree}}}\simeq 2.6\times 10^{-3}\,.
\end{equation}
With the suppression of the vectorial $Z$ coupling to leptons, we can neglect the tree level contribution $\delta C_9^{\text{tree}}$ for any $d_i\to d_j$ transition. $\abs{\delta C_{10}^{\text{tree}}}$ is $\sim 1.7\times 10^{-4}$ for $b\to d$ and reaches $0.08$ for $s\to d$ transitions, still too small to be taken into account. We neglect, consistently, all loop corrections proportional to $[\Wd]_{ij}$ with $i\neq j$. Notice that if the tree level $Z$-FCNC contributions in \refEQ{eq:treeZFCNC:00} had not been negligible, we would have had deviations from $\delta C_9=-\delta C_{10}$, since $\delta C_{9}^{\text{tree}} +\delta C_{10}^{\text{tree}}\neq 0$ controlled by $\sin^2\theta_W$.

\subsection{Lepton sector\label{sSEC:LeptonSector}}
Since the source of lepton flavour universality violation at a sizable level is, in this scenario, the box diagram in Fig. \ref{fig:Box:00} with a heavy quark $T_i$ and a heavy neutrino $N_j$ running in the loop, it is beneficial to have $\abs{\U{\mu N_j}}^2$ as large as possible and $\abs{\U{e N_j}}^2$ much smaller. With this in mind we assume for our benchmark example $\abs{\U{e N_j}}^2=0$ for all $N_j$, $\abs{\U{\mu N_j}}^2=0$ for all $N_j$ except one, $N$, for which we drop the subindex in the following, and for which $\abs{\U{\mu N}}^2$ takes a value similar to the maximum allowed by tree level flavour universality constraints \cite{Pich:2013lsa,Fernandez-Martinez:2016lgt,Escrihuela:2016ube}, following appendix \ref{APP:Univ},
\begin{equation}\label{eq:Universality:UmuN:00}
\abs{\U{\mu N}}^2\sim 10^{-3}\,.
\end{equation}
Note that with such a large value of $\abs{\U{\mu N}}^2$, to suppress $\mu\to e\gamma$ (induced by the heavy neutrino $N$ alone) one would in any case need
$\abs{\U{eN}}^2\leq 5\times 10^{-7}\sim 0$; the corresponding bound on $\abs{\U{\tau N}}^2\leq 0.24$ from $\tau\to\mu\gamma$ is irrelevant.\\
In any case, besides the previous assumption on the allowed size of the mixings, and the benchmark values of $m_N$ to be considered below, we do not refer to specific scenarios for neutrino mass generation.

\subsection{The effect in $b\to s\mu\bar\mu$\label{sSEC:bsmm}}
With the quark sector described in section \ref{sSEC:QuarkSector} and the considerations on the lepton sector in section \ref{sSEC:LeptonSector} we are finally in a position to evaluate the correction to the Wilson coefficients $C_{9,10}$ in our scenario, arising from the box diagram. For that, the relevant ingredients of our benchmark case are the generalized CKM matrix $\CKM$, the FCNC matrix $\Wu$, the heavy quark masses (essentially, only $m_{T_1}$ is relevant), the heavy neutrino admixture $\abs{\U{\mu N}}^2$, which is chosen as in \refEQ{eq:Universality:UmuN:00}, and the value of the relevant heavy neutrino mass $m_N$. To present our results, we will only have $m_N$ as a free parameter.\\ %
Taking into account that 
\begin{equation}
\frac{\Vc{T_1s}\V{T_1b}}{\Vc{ts}\V{tb}}\sim 10^{-5}\,,
\end{equation}
in order to reach a flavour dependent correction in \refEQ{eq:C9HN:00} of order $\delta C_9\sim -0.6$, it is evident that one needs $-G\sim 6\times 10^7$; now, with $m_{T_1}\sim 500$ TeV, according to Figure \ref{fig:BoxHN}, $m_N$ is required to be in the range $500$ TeV $\leq m_N\leq$ $50000$ TeV. We therefore show, for different values of $m_N$, the values at the electroweak scale of 
\begin{equation}
C_9^{\rm VLN}=C_9^{\rm VLN-Univ}+C_9^{\rm HN},\quad C_9^{\rm VLN-Univ}=C_9^{\rm SM}(3+n)+\Delta C_9^{\rm HQ}\,,
\end{equation}
where $C_9^{\rm HN}$ is dominated by the contribution with $T_1$ and $N$ (and only appears in $b\to s\mu\bar\mu$); the universal piece $C_9^{\rm VLN-Univ}$
does depend on the heavy quarks but not on the heavy neutrinos. It could in principle affect other rare processes, but its value in our benchmark scenario is
\begin{equation}
C_9^{\rm VLN-Univ}=4.42\times e^{0.001i}\,,
\end{equation}
to be compared with
\begin{equation}
C_9^{\rm SM}=4.42\,,
\end{equation}
and thus in this scenario $C_9^{\rm SM}$ for $b\to se\bar e$ takes essentially the SM value.
For the violation of lepton flavour universality, values of
\begin{equation}
\delta C_9=C_9^{\rm VLN}-C_9^{\rm SM}\sim C_9^{\rm HN}\,,
\end{equation}
are given in Table \ref{TAB:C9values}.
\begin{table}[h!tb]
\begin{center}
\begin{tabular}{|c||c|c|c|}\hline
$m_N$ & $m_{T_1}$ & $10m_{T_1}$ & $100m_{T_1}$\\ \hline
$\delta C_9$ & $-0.12\times e^{-0.025i}$ & $-0.53\times e^{-0.034i}$ & $-1.02\times e^{-0.008i}$\\ \hline
\end{tabular}
\caption{Values of $\delta C_9$\label{TAB:C9values}}
\end{center}
\end{table}

\noindent It is clear that in our scenario one can generate lepton flavour universality violations at the level given by $R_{K^{(\ast)}}$, in the context of global solutions to the existing B anomalies with $\delta C_9=-\delta C_{10}$, and with the presence of a heavy neutrino with a mass $m_N\sim 5000$ TeV, for a PMNS mixing matrix with $\abs{\U{\mu N}}^2\sim 10^{-3}$. From Table \ref{TAB:C9values} and Figure \ref{fig:BoxHN} one can see that choosing heavier neutrino masses one can go to lower $\abs{\U{\mu N}}^2$ mixings. One could similarly consider scenarios where vectorial up quark masses are lowered and the heavy neutrino mass and/or mixing rised. 

\section{Correlated Results in other Flavour Observables\label{SEC:Correlated}}
The introduction of vector-like quarks generates correlated effects in different flavour observables. We can classify them in two classes: (i) lepton flavour universal and (ii) non-universal processes. In the absence of new heavy charged leptons, any lepton flavour universality violation should come from one loop diagrams involving virtual heavy neutrinos, and thus in the first class we consider neutral meson-antimeson mixings, $q_i\to q_j\nu\bar\nu$ transitions and EW oblique corrections; in the second class we will analyse $q_i\to q_j\mu\bar\mu$.

\subsection{Lepton Flavour Universal Processes\label{sSEC:Other:LFU}}
Following \cite{Botella:2016ibj}, to which we refer for details, the corrections (modulus and phase) to the amplitude $M_{12}^{(P)}$ of the $P^0$--$\bar P^0$  mixing are, with respect to the SM prediction $M_{12}^{(P){\rm SM}}$, 
\begin{equation}\label{eq:MesonMixings:00}
\Delta(P)=\left|\frac{M_{12}^{(P)}}{M_{12}^{(P){\rm SM}}}\right|-1,\qquad \delta(P)=\frac{\arg(M_{12}^{(P)})}{\arg(M_{12}^{(P){\rm SM}})}-1.
\end{equation}
They are shown in Table \ref{TAB:MesonMixings}: they are all below current experimental uncertainties.
\begin{table}[h!tb]
\begin{center}
\begin{tabular}{|c|c|c|c|}
\cline{2-4}
\multicolumn{1}{c|}{\phantom{$\hat I$}} & $K^0$--$\bar K^0$ & $B^0_d$--$\bar B^0_d$ & $B^0_s$--$\bar B^0_s$\\ \hline
$\Delta(P)$ & $0.007$ & $0.047$ & $0.002$\\ \hline
$\delta (P)$ & $-0.088$ & $-0.032$ & $-0.055$\\ \hline
\end{tabular}
\caption{Corrections to meson mixing amplitudes, \refEQ{eq:MesonMixings:00} .\label{TAB:MesonMixings}}
\end{center}
\end{table}

\noindent For decay processes we also define the deviation with respect to SM expectations:
\begin{equation}
R(A\to B)=\frac{\Gamma(A\to B)}{[\Gamma(A\to B)]_{\rm SM}}-1\,.
\end{equation}
For our benchmark example, the different $d_i\to d_j\nu\bar\nu$ processes have
\begin{equation}
R(s\to d\nu\bar\nu)=-0.036,\quad  R(b\to d\nu\bar\nu)=-0.047,\quad  R(b\to s\nu\bar\nu)=0.002\,.
\end{equation}
Finally, for the oblique parameters \cite{Lavoura:1992np}, we have 
\begin{equation}
\Delta T=-0.005\,,\quad \Delta S=0.002\,.
\end{equation}
In summary, these results are all in agreement with existing experimental constraints.

\subsection{Lepton Flavour Universality Violating Processes\label{sSEC:Other:LFUV}}

\subsubsection{Transitions $b\to q\mu\bar\mu$\label{ssSEC:Other:bqmm}}
In this subsection we focus on the corrections to the Wilson coefficients $C_{9,10}$ (recalling that $C_9=-C_{10}$ in this scenario) for the transitions $b\to d\mu\bar\mu$ and $b\to s\mu\bar\mu$. For the benchmark example in section \ref{SEC:Example}, we consider the three different values of the heavy neutrino mass in Table \ref{TAB:C9values}; the results are shown in Table \ref{TAB:Corrections:bqmumu} (the $b\to s$ row repeats Table \ref{TAB:C9values}).
\begin{table}[h!tb]
\begin{center}
\begin{tabular}{cc|c|c|c|}\cline{3-5}
 & & $m_N=m_T$ & $m_N=10m_T$ & $m_N=100m_T$ \\ \cline{2-5}
 \phantom{$\hat I$} & \multicolumn{1}{|c|}{$C_9^{\rm VLN}$} & \multicolumn{3}{|c|}{$\delta C_9=-\delta C_{10}$}\\ \hline
\multicolumn{1}{|c|}{$b\to s\mu\bar\mu$} & $4.42\times e^{0.001i}$ & $-0.12\times e^{-0.025i}$ & $-0.53\times e^{0.0034i}$ & $-1.02\times e^{0.008i}$\\ \hline
\multicolumn{1}{|c|}{$b\to d\mu\bar\mu$} & $4.31\times e^{-0.013i}$ & $-0.22\times e^{0.056i}$ & $-0.58\times e^{-0.244i}$ & $-1.04\times e^{-0.321i}$\\ \hline
\end{tabular}
\caption{Corrections to the Wilson coefficients $C_{9,10}$ for $b\to q\mu\bar\mu$.\label{TAB:Corrections:bqmumu}}
\end{center}
\end{table}
One can see that LFU violation is quite general in $b\to d\mu\bar\mu$ once we require LFU violation in $b\to s\mu\bar\mu$ transitions. One could have expected this result except for the fact that, with different CKM matrix elements entering the process, corrections to the phases can be different in a natural way.
We can conclude that if the explanation of the $R_{K^{(\ast)}}$ anomalies is due to this model, \emph{we should expect similar effects in processes with underlying $b\to d\mu\bar\mu$ transitions}. For the benchmark example, with $m_N=10m_T$, the correction (which is already normalized to the appropriate $\Vc{tq}\V{tb}$ combination) is similar in $b\to s$ and $b\to d$ transitions; in the more suppressed $b\to d$ transition there is, however, much larger room for changes in the weak phase of the $C_9$ coefficient (this is even more apparent in the case $m_N=100m_T$ in Table \ref{TAB:Corrections:bqmumu}).
This can be readily understood with the following relation for the dominant $T_1$ mediated contributions
\begin{equation}\label{eq:bs:bd:00}
\left[\frac{\delta C_9}{C_9^{\rm SM}}\right]_{b\to s\mu\bar\mu}\sim \frac{\Vc{T_1s}}{\Vc{T_1d}}\frac{\Vc{td}}{\Vc{ts}}\left[\frac{\delta C_9}{C_9^{\rm SM}}\right]_{b\to d\mu\bar\mu}\sim e^{0.42i}\left[\frac{\delta C_9}{C_9^{\rm SM}}\right]_{b\to d\mu\bar\mu}\,,
\end{equation}
which shows that the corrections in $b\to s\mu\bar\mu$ and $b\to d\mu\bar\mu$ transitions have similar size. The phase is not universal in Table \ref{TAB:Corrections:bqmumu} since the flavour universal contribution can have a phase modified with respect to the SM and, with different contributions, the final phase of the correction may differ. 

\subsubsection{Transitions $s\to d\mu\bar\mu$\label{ssSEC:Other:dsmm}}
The transitions $s\to d\mu\bar\mu$ have to be analysed separately. The analog of \refEQ{eq:bs:bd:00} would read
\begin{equation}\label{eq:bs:ds:00}
\left[\frac{\delta C_9}{C_9^{\rm SM}}\right]_{s\to d\mu\bar\mu}\sim \frac{\V{T_1d}}{\V{T_1b}}\frac{\V{tb}}{\V{td}}\left[\frac{\delta C_9}{C_9^{\rm SM}}\right]_{b\to s\mu\bar\mu}\sim 25\times e^{0.42i}\left[\frac{\delta C_9}{C_9^{\rm SM}}\right]_{b\to s\mu\bar\mu}\,,
\end{equation}
which certainly hints at much more important effects than the ones in $b\to q\mu\bar\mu$ transitions.
Nevertheless, it is well known that in general it is not trivial to observe short distance contributions such as the ones controlled by $C_9$ and $C_{10}$ (also labelled in the context of kaon decays as $C_{7V}$ and $C_{7A}$) due to the low $q^2$ available. These transitions are dominated by long distance one photon contributions: we will show that LFUV contributions in $s\to d\mu\bar\mu$ at the level of $\delta C_9\sim -3e^{0.42i}C_9^{\rm SM}$, or even larger, are still allowed. We follow reference \cite{Crivellin:2016vjc}, but using the normalization of the Wilson coefficients in \refEQS{eq:Leff:bs:00} for the kaon case, i.e. as in \eqref{eq:Leff:bs:00} with $\V{tb}\Vc{ts}\mapsto \V{ts}\Vc{td}$; the LFUV correction $\delta C_9$ verifies
\begin{equation}\label{eq:KaonC9:00}
\V{td}\Vc{ts}\,\delta C_9^{s\to d\mu\bar\mu}=-\frac{a_+^{\mu\mu}-a_+^{ee}}{\sqrt 2}
\end{equation}
with $a_+^{\ell\ell}$ parameters appearing in the $K^\pm\to\pi^\pm\ell\bar\ell$ decay spectra. According to \cite{Crivellin:2016vjc} (and references therein) we use the experimental input $a_+^{\mu\mu}-a_+^{ee}=0.007\pm 0.040$, and \refEQ{eq:KaonC9:00} gives
\begin{equation}\label{eq:KaonC9:01}
\left[\V{td}\Vc{ts}\,\delta C_9^{[s\to d\mu\bar\mu]}\right]_{\rm Exp.}=-0.005\pm 0.028\,,
\end{equation}
to be compared with the prediction for the three different $m_N$ values:
\begin{equation}\label{eq:KaonC9:Examples:00}
\V{td}\Vc{ts}\,\delta C_9^{[s\to d\mu\bar\mu]}\simeq 
\left\{\begin{matrix}0.001 &\text{ for }&m_N=m_T,\\ 0.004 &\text{ for }&m_N=10m_T,\\ 0.008 &\text{ for }&m_N=100m_T.\end{matrix}\right.
\end{equation}
Imaginary parts in \refEQ{eq:KaonC9:Examples:00} are much smaller and have been omitted. This explains that the effects of this NP scenario are far from being relevant in $K^\pm\to\pi^\pm\ell\bar\ell$ processes with the current precision level. Nevertheless, the large factor in \refEQ{eq:bs:bd:00}, $\sim 25$, implies that one could expect sizable modifications in rare kaon decays where short distance effects are important. This is the case of the decay $K_L\to\pi^0\ell\bar\ell$, which is dominated by CP violating pieces. Following \cite{Crivellin:2016vjc,Buchalla:2003sj}, one can write
\begin{equation}
\left.\text{BR}(K_L\to\pi^0 \mu\bar \mu)\right|_{\rm CPV}=10^{-12}\left(3.5\abs{a_s^\mu}^2+1.5\abs{a_s^\mu}\left[\frac{\im{\V{td}\Vc{ts}}}{10^{-4}}\right]+1.1\left[\frac{\im{\V{td}\Vc{ts}}}{10^{-4}}\right]^2\right),\label{eq:KLpimm:00}
\end{equation}
where $\abs{a_s^\mu}=1.54\pm 0.40$. 
The direct CP violating terms coming from the SM and the LFUV contributions to $K_L\to\pi^0\mu\bar\mu$ are proportional to
\begin{equation}\label{eq:KLpill:01}
\im{\V{td}\Vc{ts}(C_9^{\rm SM}+\delta C_9)}=C_9^{\rm SM}\,\im{\V{td}\Vc{ts}\left(1+\left[\frac{\delta C_9}{C_9^{\rm SM}}\right]_{d\to s\mu\bar\mu}\right)}\,,
\end{equation}
which can have the following values:
\begin{equation}\label{eq:KLpimumu:00}
\V{td}\Vc{ts}\left(1+\left[\frac{\delta C_9}{C_9^{\rm SM}}\right]_{d\to s\mu\bar\mu}\right)=
\left\{
\begin{matrix}
(-1.00+1.30i)10^{-4}&\text{ for }&m_N=m_{T_1},\\
(6.45+1.60i)10^{-4}&\text{ for }&m_N=10m_{T_1},\\
(15.8+1.97i)10^{-4}&\text{ for }&m_N=100m_{T_1}.
\end{matrix}
\right.
\end{equation}
The effects in $K_L\to\pi^0\mu\bar\mu$ can be estimated substituting $\im{\V{td}\Vc{ts}}$ in \refEQ{eq:KLpimm:00} with \refEQ{eq:KLpill:01}; only the imaginary parts are relevant\footnote{It is clear that the observables should be invariant under rephasings of the CKM matrix while \refEQ{eq:KLpimm:00} is not: it is to be understood that $\im{\V{td}\Vc{ts}}$ is computed in a phase convention where $\V{cd}\Vc{cs}$ is real.}. For the $3\times 3$ unitary CKM matrix of the SM, $[\V{td}\Vc{ts}]_{\rm SM}=(-3.1548+1.3652i)10^{-4}$, and thus, in \refEQ{eq:KLpimm:00}
\begin{multline}\label{eq:KLpill:02}
\left[\frac{\im{\V{td}\Vc{ts}}}{10^{-4}}\right]\sim 1.365\text{ {\rm(SM)} }\\
\to\ 10^{4} \im{ \V{td}\Vc{ts}\left(1+\left[\frac{\delta C_9}{C_9^{\rm SM}}\right]_{d\to s\mu\bar\mu}}\right)=
\left\{\begin{matrix}
1.30&\text{ for }&m_N=m_{T_1},\\
1.60&\text{ for }&m_N=10m_{T_1},\\
1.97&\text{ for }&m_N=100m_{T_1}.
\end{matrix}\right.
\end{multline}
The effect can be relevant, but testing it requires improvements in both the theoretical knowledge of the long distance contribution and in the foreseen experimental measurements \cite{Komatsubara:2012pn}.
Since $C_9=-C_{10}$, one must also analyse the effect of $C_{10}$ on $K_L\to\mu\bar\mu$. The dominant contribution, however, comes from the $2\gamma$ intermediate state, obscuring again the analysis of the expected effect. In this CP conserving process, what matters in the short distance contribution is the real part, 
and thus, similarly to \refEQ{eq:KLpill:02} and using again \refEQ{eq:KLpimumu:00},
\begin{multline}\label{eq:KLll:01}
\left[\frac{\re{\V{td}\Vc{ts}}}{10^{-4}}\right]\sim -3.15\text{ {\rm (SM)} }\\
\to\ 10^{4} \re{ \V{td}\Vc{ts}\left(1+\left[\frac{\delta C_9}{C_9^{\rm SM}}\right]_{d\to s\mu\bar\mu}}\right)=
\left\{\begin{matrix}
-1.00&\text{ for }&m_N=m_{T_1},\\
6.45&\text{ for }&m_N=10m_{T_1},\\
15.8&\text{ for }&m_N=100m_{T_1}.
\end{matrix}\right.
\end{multline}
Substituting \refEQ{eq:KLll:01} in the short distance contribution to $K_L\to\mu\bar\mu$ gives \cite{Gorbahn:2006bm,Cirigliano:2011ny}
\begin{equation}\label{eq:KLll:02}
\left.\text{BR}(K_L\to\mu\bar\mu)\right|_{SD}=10^{-9}\times 
\left\{\begin{matrix}
0.15&\text{ for }&m_N=m_{T_1},\\
2.03&\text{ for }&m_N=10m_{T_1},\\
13.6&\text{ for }&m_N=100m_{T_1}.
\end{matrix}\right.
\end{equation}
A conservative upper bound  \cite{Cirigliano:2011ny} is $\left.\text{BR}(K_L\to\mu\bar\mu)\right|_{SD}\leq 2.5\times 10^{-9}$, which clearly disfavours or excludes the example for $m_N=100m_{T_1}$.
We can conclude that rare kaon decay processes sensitive to short distance contributions will be highly relevant to prove or disprove the proposed scenario. Specifically, $K_L\to\pi^0\mu\bar\mu$ and $K_L\to\mu\bar\mu$ show correlated effects with the B anomalies, including $R_{K^{(\ast)}}$. From this analysis of the kaon sector, we can conclude that it is the benchmark example with $m_N\simeq 10 m_{T_1}$ that fulfills all the considered theoretical and experimental constraints in a natural way.


\clearpage
\section{Conclusions\label{APP:Conclusions}}
We have shown that the existence of heavy neutrinos $N_j$ leading to an entry in the PMNS matrix that couples the neutrino $N$ with muons, together with the existence of vectorial quarks, can explain the so-called B anomalies in $b\to s\ell\bar\ell$ transitions in a scenario where $\delta C_9\simeq -\delta C_{10}$. The New Physics contribution to $C_{9,10}$ explaining $R_{K^{(\ast)}}$ originates from the standard box diagram, but with heavy neutrinos and new vectorial up quarks running inside the loop. In general, without additional vectorial up quarks, the top quark contribution is too small to account for the anomalies. To allow for deviations from Lepton Flavour Universality and simultaneously avoid too large FCNC effects, the singlet vectorial quarks  should be heavy enough, specially if we take into account that we need them to couple with ordinary quarks through charged current interactions. To avoid suppressions in the box diagram or, conversely, to maximize the effects, the heavy neutrino should have masses similar to the vectorial quarks, or larger.
We have introduced vectorial quarks that obey special flavour symmetries for which (i) the CKM matrix would be the identity and (ii) the four lighter quarks would be massless if the couplings to the new vector-like quarks were absent; in their presence, both the deviation of CKM from the identity and the masses of the lighter quarks originate from the mixing among the standard and the vector-like quarks. 
It is well known that in general this kind of quarks generate, at the loop level, very large LFU FCNC effects, that can only be suppressed if they decouple. Consequently, the masses of the vectorial quarks should be heavy enough: in our example we have considered vectorial quarks with masses of order 10 TeV and 500 TeV for, respectively, down and up-type quarks. With such heavy masses, all LFU effects are in general highly suppressed, but this is not the case for LFUV effects in transitions other than $b\to s\mu\bar\mu$ when the $R_{K^{(\ast)}}$ anomalies are accounted for.  In particular we find similar LFUV effects in $b\to d\mu\bar\mu$ transitions that should be tested. Even larger effects in the short distance contributions to $d\to s\mu\bar\mu$ processes are also found: $K_L\to\mu\bar\mu$ is now the most relevant constraint, but $K_L\to\pi^0\mu\bar\mu$ will also be relevant in the future. In general, we do not have effects in $q_i\to q_j\nu\bar\nu$ transitions because we have not introduced heavy charged leptons.\\ 
In summary, we have shown that a simple extension of the Standard Model, involving the introduction of vector-like quarks and heavy neutrinos can provide a possible explanation for the B anomalies in $b\to s\ell\bar\ell$ transitions, in a framework which also provides a rationale for the almost 
decoupling of the heavier third quark generation with $\abs{\V{tb}}^2=1-1.6\times 10^{-3}$.

\clearpage
\section*{Acknowledgments}
The authors acknowledge Francisco del Águila, Quim Matías, Mariano Quirós, Arcadi Santamaría, Joaquim Silva-Marcos and Óscar Vives for interesting comments or discussions.
This work is partially supported by Spanish MINECO under grant FPA2015-68318-R and by the Severo Ochoa Excellence Center Project SEV-2014-0398, by Generalitat Valenciana under grant GVPROMETEOII 2014-049 and by Funda\c{c}\~ao para a Ci\^encia e a Tecnologia (FCT, Portugal) through the projects CERN/FIS-NUC/0010/2015 and CFTP-FCT Unit 777 (UID/FIS/00777/2013) which are partially funded through POCTI (FEDER), COMPETE, QREN and EU. 
MN acknowledges support from FCT through postdoctoral grant SFRH/BPD/112999/2015.


\appendix

\section{Example\label{APP:EX}}
The example of section \ref{sSEC:QuarkSector} is completely defined in terms of the mass matrices in \refEQS{eq:QLMass:00}--\eqref{eq:MassBlocks:00}:
\begin{equation}
\wMD\equiv
\begin{pmatrix}\mqd & \omqd \\ \Xqd & \Mqd\end{pmatrix}.\qquad
\wMU\equiv
\begin{pmatrix}\mqu & \omqu \\ \Xqu & \Mqu\end{pmatrix}\,.\tag{\ref{eq:MassBlocks:00}}
\end{equation}
They have the following structure, with all entries in GeV.
\begin{equation}\label{eq:MassM:m:00}
\mqd=\begin{pmatrix}0 & 0 & 0\\ 0 & 0 & 0\\ 0 & 0 & -0.6-1.2i\end{pmatrix}\,\qquad
\mqu=\begin{pmatrix}0 & 0 & 0\\	0 & 0 &	0\\ 0 & 0 & -70.8131+192.531i\end{pmatrix},
\end{equation}
notice that without mixing with the new heavy quarks, \refEQ{eq:MassM:m:00} gives at leading order $\CKM=\mathbf{1}$ and massless light quarks with $m_b$ and $m_t$ close to the actual values. The upper right blocks read
\begin{align}\label{eq:MassM:om:00}
\omqd&=\begin{pmatrix}0.2542 & 0 & 0\\ 3.2994+4.6982i & 11.067 & 0\\ -0.3877-0.8229i & 55.649-2.3949i & 1.7419\end{pmatrix},\\
\omqu&=\begin{pmatrix}2.5148 & 0 & 0\\ 93.5524-255.32i & 20.8841 & 0\\ -122.292+332.496i & 107.905-31.0875i & 2.0922\end{pmatrix},
\end{align}
with entries not larger than the electroweak scale. The $\Delta I=0$ blocks of the mass matrices are
\begin{equation}\label{eq:MassM:M:00}
\Mqd=\begin{pmatrix}1 & 0 & 0\\ 0 & 1.1 & 0\\ 0 & 0 & 1.2\end{pmatrix}\times 10^4,\qquad
\Mqu=\begin{pmatrix}5 & 0 & 0\\ 0 & 5 & 0\\ 0 & 0 & 5\end{pmatrix}\times 10^5,
\end{equation}
and
\begin{equation}\label{eq:MassM:X:00}
\Xqd=\begin{pmatrix}0 & 0 & 500\\ 0 & 500 & 0\\ 500 & 0 & 0\end{pmatrix},\qquad
\Xqu=\begin{pmatrix}0 & 0 & 5\\ 0 & 5.625 & 0\\ 6.25 & 0 & 0\end{pmatrix}\times 10^4\,.
\end{equation}
The scale of these matrices is larger than the electroweak scale: the off-diagonal submatrices are smaller than the diagonal submatrices, the former being responsible of giving masses to the light quarks and generating CKM mixings, and the later being responsible of the heavy quark masses. Following the usual bi-diagonalisation of $\wMD\wMDd$ and $\wMU\wMUd$, we obtain the mass spectrum (at the $M_Z$ scale) in \refEQ{eq:QMasses:00} and the diagonalizing matrices that give $\CKM$, $\Wu$ and $\Wd$ shown in section \ref{SEC:Example}, \refEQS{eq:CKM:Example:00}--\eqref{eq:Wd:Example:00}.\\ 
Finally, following the ideas of reference \cite{Botella:2016ibj}, the zero entries of the matrices $\mqd$, $\mqu$, $\omqd$ and $\omqu$ should be enforced by an abelian symmetry, under which a field $\Psi$ transforms as
\begin{equation}
\Psi\mapsto e^{i\tau\,Q(\Psi)}\Psi\,.
\end{equation}
An example that meets these requirements for the matrices above is a model with four Higgs doublets $\phi_i$ and the following charge assignments for the fields in the model:
\begin{equation}
Q\begin{pmatrix} u^0_{R1}\\ u^0_{R2}\\ u^0_{R3}\\ u^0_{R4}\\ u^0_{R5}\\ u^0_{R6}\end{pmatrix}=
\begin{pmatrix} -1\\ -1\\ 6\\ 4\\ 5\\ 6\end{pmatrix}\,,\qquad
Q\begin{pmatrix} d^0_{R1}\\ d^0_{R2}\\ d^0_{R3}\\ d^0_{R4}\\ d^0_{R5}\\ d^0_{R6}\end{pmatrix}=
\begin{pmatrix} 5\\ 4\\ 3\\ 0\\ 2\\ 3\end{pmatrix}\,,
\end{equation}

\begin{equation}
Q\begin{pmatrix} \phi_{1}\\ \phi_{2}\\ \phi_{3}\\ \phi_{4}\end{pmatrix}=
\begin{pmatrix} 1\\ 0\\ 2\\ 3\end{pmatrix}\,,\qquad
Q\begin{pmatrix} Q^0_{L1}\\ Q^0_{L2}\\ Q^0_{L3}\end{pmatrix}=
\begin{pmatrix} 1\\ 2\\ 3\end{pmatrix}\,.
\end{equation}
With these assignments it is easy to reproduce the flavor structure of the $\Delta I=1/2$ mass matrices. If the symmetry in the $\Delta I=0$ mass matrices is broken, it amounts to soft breaking without spoiling the nice features of the model. For this reason it is not necessary to worry about the charge assignments of the left handed components of the singlet vectorial quarks: $D_{L_{1}}^{0}$, $D_{L_{2}}^{0}$, $D_{L_{3}}^{0}$, $T_{L_{1}}^{0}$, $T_{L_{2}}^{0}$ and $T_{L_{3}}^{0}$.
\section{Universality\label{APP:Univ}}
\begin{figure}[h!bt]
\begin{center}
\subfigure[\label{fig:Universality:a}]{\includegraphics[width=0.32\textwidth]{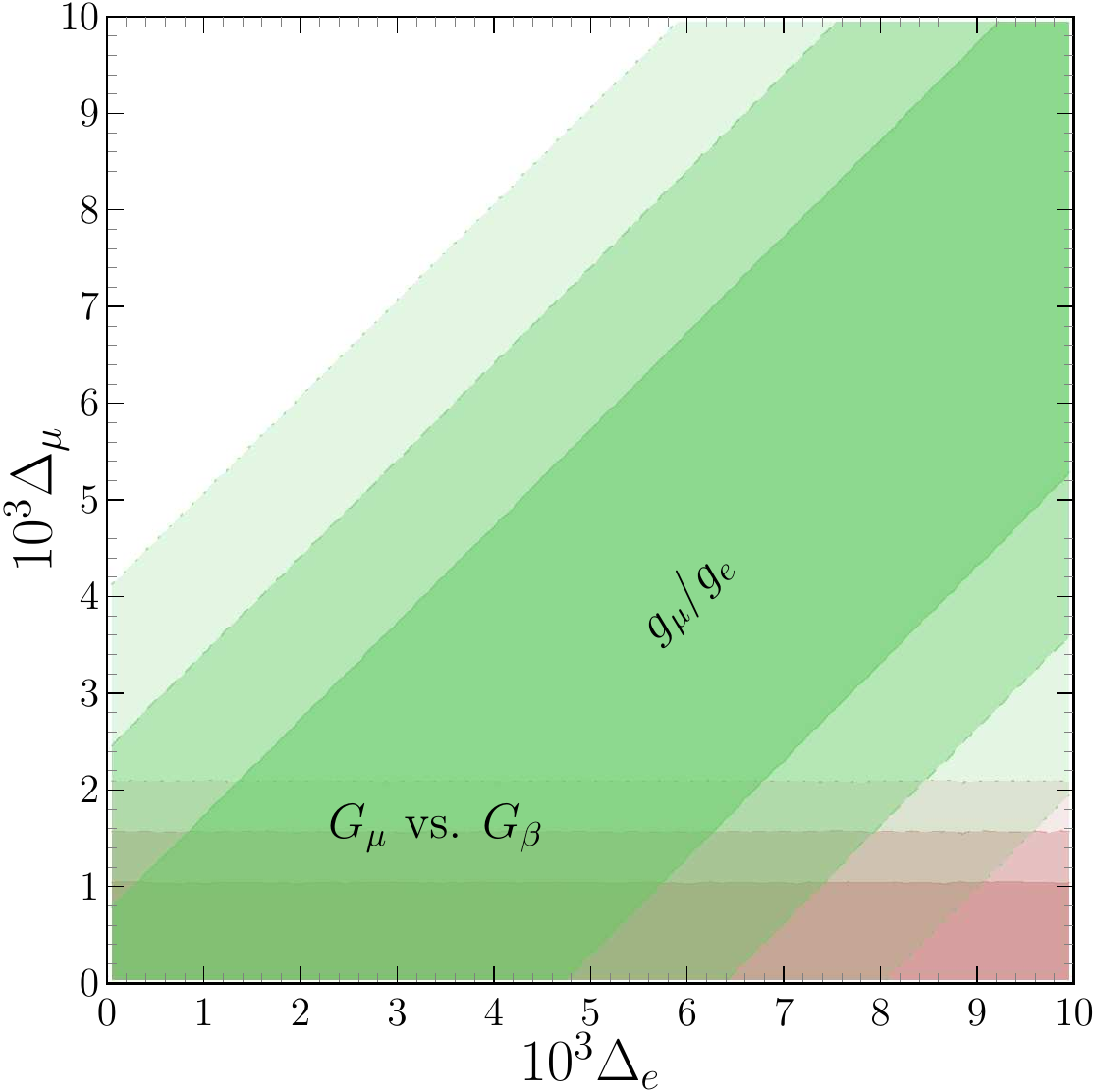}}\qquad
\subfigure[\label{fig:Universality:b}]{\includegraphics[width=0.32\textwidth]{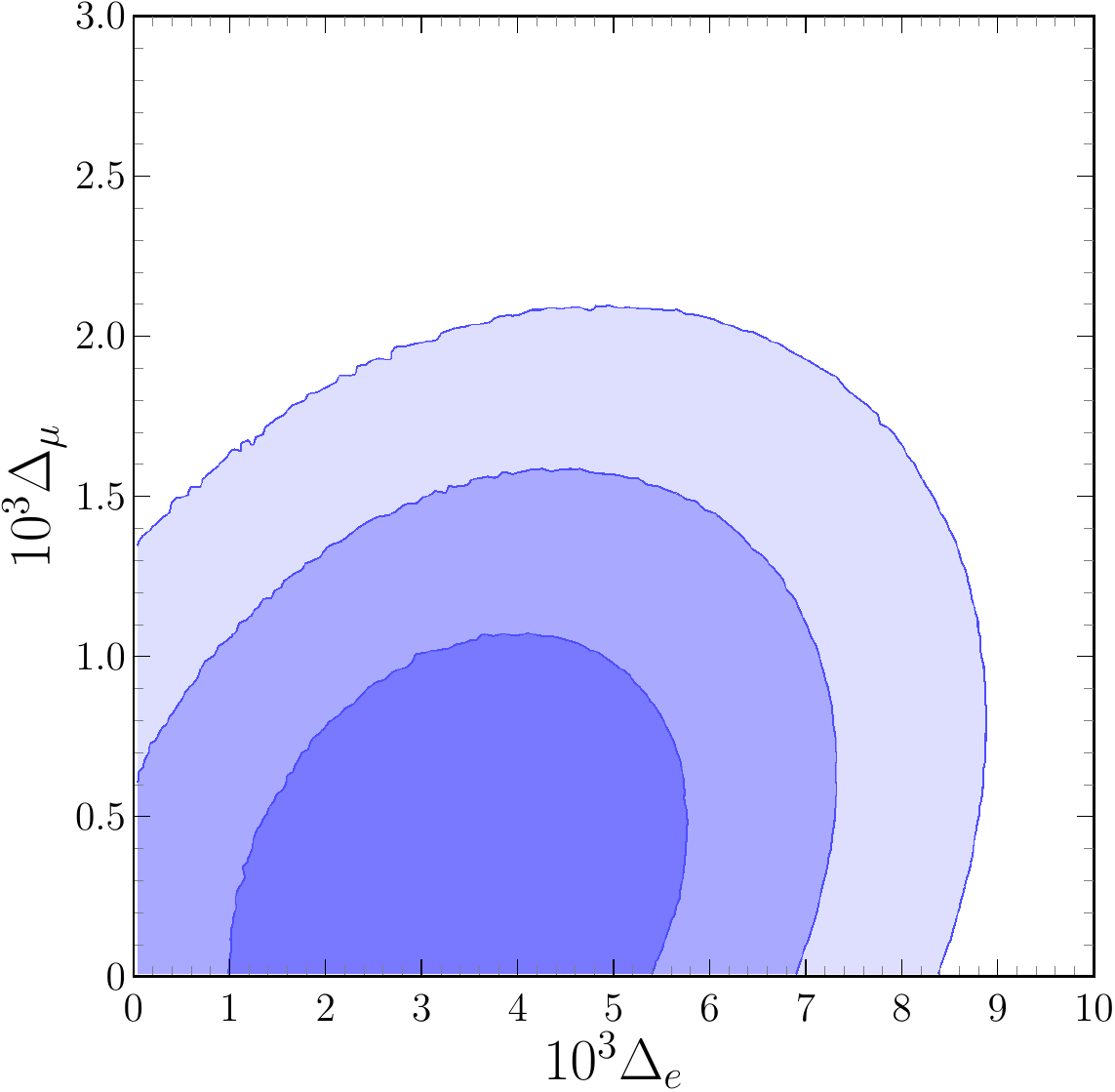}}
\caption{Allowed $\Delta_\mu$ vs. $\Delta_e$ regions (68, 95 and 99\% C.L.) by universality constraints.\label{fig:Universality}}
\end{center}
\end{figure}

Deviations of the PMNS mixing matrix from $3\times 3$ unitarity would appear as violations of universality of weak interactions in some low energy processes. The (incoherent) sum over final state neutrinos gives for instance
\begin{align}\label{eq:PMNS:3x3devs:CC}
&\Gamma(W\to\ell\nu)\propto\sum_{j=1}^3\abs{U_{\ell j}}^2\equiv 1-\Delta_\ell\,,
\\ \label{eq:PMNS:3x3devs:CC2}
&\Gamma(\ell_1\to\ell_2\nu\bar\nu)\propto\sum_{j,k=1}^3\abs{U_{\ell_1 j}}^2\abs{U_{\ell_2 k}}^2\equiv (1-\Delta_{\ell_1})(1-\Delta_{\ell_2})\,,
\end{align}
in the charged current coupling of leptons $\ell=$ $e$, $\mu$, $\tau$. For a non $3\times 3$ unitary PMNS involving heavy enough neutrinos (i.e. which are not included in the sum over final states), the different $\Delta_\ell$ in \refEQS{eq:PMNS:3x3devs:CC}--\eqref{eq:PMNS:3x3devs:CC2} parametrise those deviations. It is common to translate this effect into ``non-universal'' weak coupling constants for the different leptons
\begin{equation}\label{eq:PMNS:3x3devs:NonU}
g\mapsto g_\ell=g\sqrt{1-\Delta_\ell}\,.
\end{equation}
Constraints on the different $g_\ell$ can then be obtained from a number of observables.
\begin{itemize}
\item Comparison of the Fermi constant measured in muon decay, $G_\mu$, and in (nuclear) beta decays, $G_\beta$: while the former involves both $g\mapsto g_\mu$ and $g\mapsto g_e$, the latter involves $g\mapsto g_e$ and the CKM matrix element $\V{ud}$. Including information on $\V{us}$ would turn this comparison into a simultaneous test of unitarity of both PMNS and CKM. The usual extraction of $\V{us}$ \cite{Olive:2016xmw} involves the combined use of kaon decays into $\mu\nu$ and into $e\nu$ final states, which we better avoid in this context. However, on that respect, $\abs{\V{us}/\V{ud}}$ can be extracted \cite{Olive:2016xmw} from $K\to\mu\nu(\gamma)$ and $\pi\to\mu\nu(\gamma)$, that is free from $\Delta_e$ and $\Delta_\mu$. Assuming universality and no significant unitarity deviations in CKM (one can in fact neglect $\abs{\V{ub}}^2$ to start with), $G_\mu$ vs. $G_\beta$ gives
\begin{equation}\label{eq:GFermi:00}
g_\mu^2\,g_e^2=g_e^2g^2\abs{\V{ud}}^2\left(1+\frac{\abs{\V{us}}^2}{\abs{\V{ud}}^2}\right)\\
\Leftrightarrow\,(1-\Delta_\mu)=\abs{\V{ud}}^2\left(1+\frac{\abs{\V{us}}^2}{\abs{\V{ud}}^2}\right)\,.
\end{equation}
Equation \eqref{eq:GFermi:00} provides a constraint on the absolute size of $\Delta_\mu$, especially important since the remaining constraints only restrict ratios $g_{\ell_1}/g_{\ell_2}$.
\item $g_\mu/g_e$ is constrained by (a) semileptonic decays $P\to\mu$ vs. $P\to e$ ($P=K,\pi$), and $K\to\pi\mu$ vs. $K\to\pi e$, and (b) leptonic decays $\tau\to\mu$ vs. $\tau\to e$.
\item $g_\tau/g_\mu$ is constrained by (a) semileptonic decays $\tau\to P$ vs. $P\to\mu$ ($P=K,\pi$), and (b) leptonic decays $\tau\to e$ vs. $\mu\to e$.
\item $g_\tau/g_e$ is constrained by leptonic decays $\tau\to\mu$ vs. $\mu\to e$.
\end{itemize}
With \cite{Olive:2016xmw} $\abs{\V{ud}}=0.97417\pm 0.00021$, $\abs{\V{ud}}=0.2254\pm 0.0008$, \refEQ{eq:GFermi:00} gives a simple estimate
$\Delta_\mu=\left(5.1\begin{smallmatrix}+ 4.3\\ - 3.0\end{smallmatrix}\right)\times 10^{-4}$.\\ 
For the constraints on $g_{\ell_1}/g_{\ell_2}$, we refer to Table 2 in \cite{Pich:2013lsa}; it is to be noticed that $g_\tau$ related constraints play no significant role and a simple analysis in terms of $\Delta_e$ and $\Delta_\mu$ can be easily done: figure \ref{fig:Universality:a} shows the regions in the $(\Delta_e,\Delta_\mu)$ plane allowed by the separate constraints (i) $G_\mu$ vs. $G_\beta$ and (ii) $g_\mu/g_e$. 
The combined allowed region is shown in figure \ref{fig:Universality:b}. With these constraints, the largest value of $\Delta_\mu$, that is, the largest $\abs{\U{\mu j}}^2$ that can be considered, is $\abs{\U{\mu j}}^2\simeq 1.8\times 10^{-3}$ in the 3 $\sigma$ region. Forcing $\Delta_e=0$, one can consider values of $\abs{\U{\mu j}}^2$ up to $\simeq 1.4\times 10^{-3}$ within the 3 $\sigma$ region.
Notice that in scenarios with heavy neutrinos and deviations from $3\times 3$ unitarity of the PMNS matrix, one may also expect effects in electroweak precision observables and in lepton flavour violating processes like $\ell_2\to\ell_1\gamma$ or $\mu$--$e$ conversion; for detailed analyses with such ingredients, see for example \cite{Fernandez-Martinez:2016lgt,Escrihuela:2016ube}. However, these effects are rather model dependent, and for that reason we consider the simple bounds derived above.

\newpage

\providecommand{\href}[2]{#2}\begingroup\raggedright\endgroup

\end{document}